\documentclass[prl,aps,showpacs,groupedaddress,twocolumn]{revtex4}
\pdfoutput=1
\usepackage{amsmath,amsfonts,amssymb}
\usepackage{wrapfig}
\usepackage{graphicx}
\usepackage{bbm}
\usepackage{graphics}

\def \beq {\begin{equation}}
\def \eeq {\end{equation}}
\def \ba {\begin{eqnarray}}
\def \ea {\end{eqnarray}}

\newcommand{\mean}[1]{\langle#1\rangle}

\renewcommand{\section}[1]{\emph{#1}--}

\begin{document}

\title{Laser cooling and optical detection of excitations in a LC  
electrical circuit}
\author{J. M. Taylor$^{1}$, A. S. S{\o}rensen$^{2}$, C. M. Marcus$^{3}$, E. S. Polzik$^{2}$}
\date{\today}
\affiliation{$^1$Joint Quantum Institute/NIST, College Park, Maryland\\
$^2$QUANTOP, Niels Bohr Institute, University of Copenhagen, Denmark \\
$^3$Department of Physics, Harvard University, Cambridge, MA}
\begin{abstract} 
We explore a method for laser cooling and optical detection of excitations in a 
 LC electrical circuit. Our approach uses a nanomechanical oscillator as a transducer between optical and electronic excitations. An experimentally feasible system with the oscillator capacitively coupled to the LC and at the same time interacting with light via an optomechanical force is shown to provide strong electro-mechanical coupling.  Conditions for  
improved sensitivity and quantum limited readout of 
electrical signals with such an ``optical loud speaker'' are outlined.
\end{abstract}
\maketitle

Cooling  plays an essential roles in most areas of physics, in part because it reduces detrimental thermal fluctuations.  For sensing application, where  
thermal fluctuations may hide the small signals one is trying to  
measure, strong coupling of mechanical and electrical oscillators to systems in  
a pure quantum state, such as light  or polarized atomic ensembles,  
opens up new possibilities for quantum sensing of fields and forces~\cite{braginsky92}.   
This in principle allows for enhanced sensitivity of the oscillators, where readout of their state is limited only by
quantum fluctuations. In recent years, dramatic advances in  
optomechanical coupling and cooling of high quality-factor (Q) mechanical systems have  
been made \cite{Kippenberg,Aspelmeyer,Tombesi,Vahala,Karrai,Girvin}. 

In this  
Letter, we propose to extend laser cooling of mechanical objects  to  
electrical circuits. By coupling a high-Q inductor-capacitor resonator (LC)  to a near-resonant nanomechanical membrane~\cite{harrisXX,Kimble} in the radio frequency (rf) domain, the electrical circuit can be  
effectively cooled by the cold mechanical system. Since  
such electrical circuits are used in a wide variety of settings, the  
reduction of thermal fluctuations in these system will likely find numerous  
applications.  We also show that the  
cooling techniques explored here allow for optical readout of  
electrical signals in the circuit. Since light fields are  
routinely measured with quantum limited precision this allows for high  
sensitivity broadband detection of weak electric signals.
Moreover, light and atomic ensembles can behave as oscillators with a negative mass and thus provide the means to measure fields and forces beyond the standard quantum limit using the power of entanglement 
\cite{Wasilewski,Hammerer,Caves}.  Finally, our coupling occurs at the level of individual radio frequency photons; thus, for systems with sufficiently low thermal load, our approach provides a versatile interface for quantum information, allowing for the reliable transfer of quantum states from radio frequency to optical domains and back.  

\begin{figure}
\includegraphics[width=3.0in]{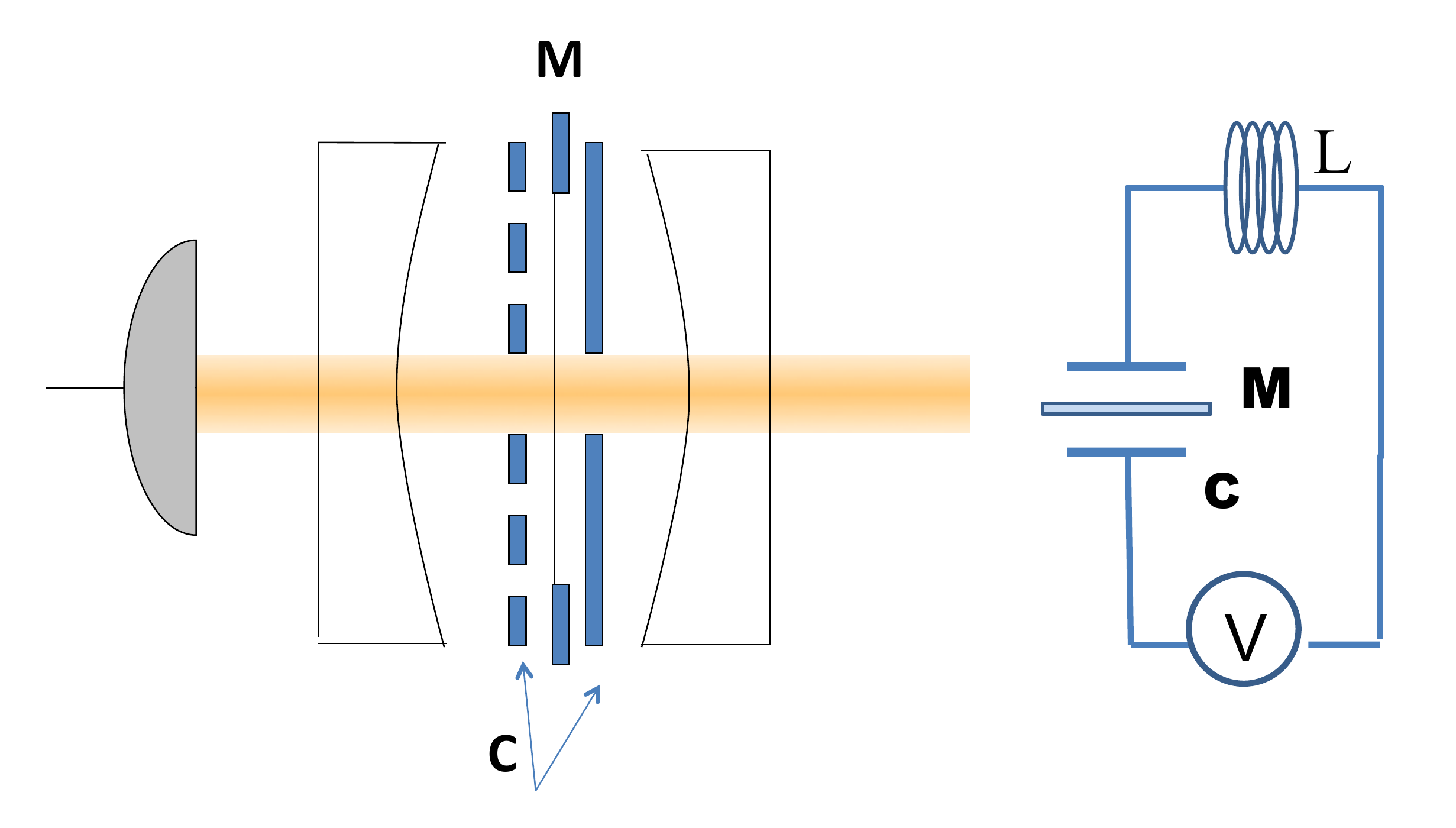}
\caption{
(left) Schematic of a Fabry-Perot cavity with a nanomechanical membrane inserted in the waist of the cavity; the membrane, in turn, is part of a parametric capacitor.  (right) Equivalent circuit with a dc voltage bias describing the coupled electromechanical system.  
\label{f:setup}}
\end{figure}

The key idea in this work is to achieve sufficient coupling between a nanomechanical membrane and a high-Q LC circuit to induce normal mode splitting~\cite{simmonds10}, 
 and then observe the electrical excitations via opto-mechanical coupling between the membrane and a high-Q optical cavity.  Our suggested method is insertion of the membrane into the fringing field of a capacitor~\cite{silvan}, as shown in Fig.~\ref{f:setup}, such that the capacitance depends upon the displacement of the membrane. 
We demonstrate that  for reasonable component parameters, when combined with a voltage bias and an inductive component to make a resonant electrical circuit near the frequency of a mechanical resonance $\omega_m$, the coupling $g$ between radio-frequency (rf) photons and the membrane phonons can become sufficiently large to induce normal mode splitting, where the resonant response of the system comprises combined electro-mechancial excitations.  
 
A model Hamiltonian describing the coupled electromechanical system (Fig.~\ref{f:setup}) in the limit of well separated, high-Q resonant electrical and mechanical modes is
\beq
H = \frac{\phi^2}{2 L} + \frac{q^2}{2C} + \frac{p_m^2}{2 m} + \frac{m \omega_m^2 x_m^2}{2} + \frac{g}{q_0 x_0} (q x_m)\ .
\eeq
Here $\hbar =1$, 
$q_0 = 1/\sqrt{2 L \omega_0}, x_0 = 1/\sqrt{2 m \omega_m}$, and $\omega_0 = \sqrt{1/LC}$. The flux $\phi=L dq/dt$ and the membrane momentum  $p_m=m dx_m/dt$ are the canonical momenta conjugate to the charge $q$ and position $x_m$. When the two modes are brought into resonance $\omega_m= \omega_0 = \omega$, the natural canonical variables become normal mode solutions $Y_\pm,P_\pm$ with $x_m = (Y_+ + Y_-)/\sqrt{2 m}$; $q = (Y_+ - Y_-)/\sqrt{2 L}$; $p_m = (Y_+ + Y_-) \sqrt{m /2}$; $\phi = (Y_+ - Y_-) \sqrt{L /2}$, with
frequencies $\omega_{\pm} = \omega \sqrt{1 \pm g/\omega}$  
for $g<\omega$ as considered below.

This coupled-mode system could be inserted into an optical cavity (Fig.~\ref{f:setup}), whose mode couples to the membrane position $x_m$ via  
radiation pressure~\cite{harrisXX}. Both normal modes interact with the cavity mode as $x_m \propto Y_+ + Y_-$; thus, radiation pressure-based detection can be applied independently to each normal mode when the cavity linewidth $\kappa$ is narrow enough to resolve the normal mode splitting $\omega_+ - \omega_-$.  Alternatively, both normal modes can be observed simultaneously when $|\omega_-| > \kappa > \omega_+ - \omega_-$.  
In both cases the optical system can also cool the combined electro-mechanical system. 
In essence this cooling is achieved because the membrane-cavity system acts as a transducer up-converting excitations in the LC circuits to optical frequencies. This means that the LC circuit will equilibrate with the optical modes which are in the quantum mechanical ground state even at room temperature. At the same time the interface between rf excitations in the LC circuits and optical photons also allows for detection of rf-electric signals by optical measurement. Since optical measurement can by quantum noise limited this opens up new possibilities for detection weak electrical signals,  as we outline below.

\begin{figure}[htbp]
\begin{center}
\includegraphics[width=8.5cm]{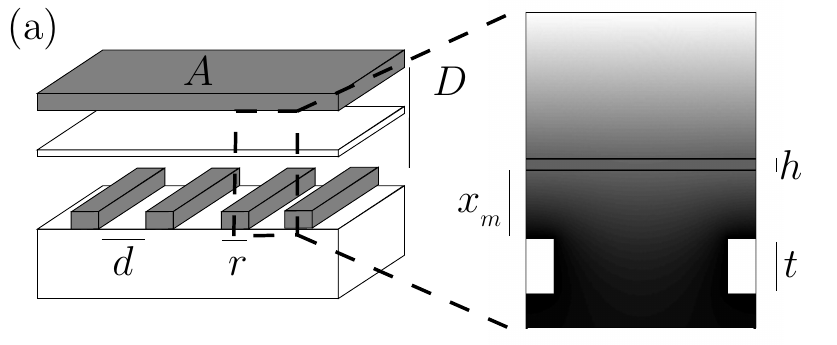}
\includegraphics[width=8.5cm]{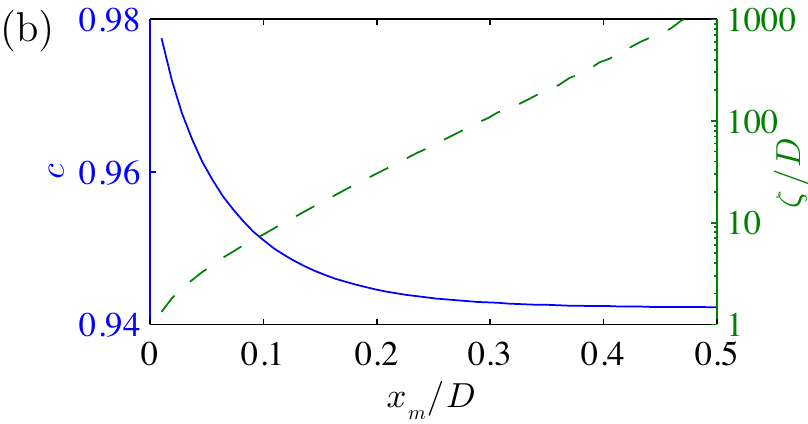}
\caption{Coupling of capacitor and membrane. (a) Schematic of the considered setup (left). A capacitor is made of  a plate of area $A$ separated from a set of wires by a distance $D$. The wires have thickness $t$ and width $r$ and are separated by a distance $d$. A membrane of thickness $h$ is separated from the wires by a distance $x_m$.  To find the capacitance we perform FEM simulations over a cross section of the capacitor (right). The gray scale indicates the simulated electric potential with $r=D/4$, $d=3D/4$, $t=D/4$, $h=D/20$, and $\epsilon=7.6$. (b) Ratio of capacitance with the membrane to a parallel plate geometry without membrane, $c$ (full curve, left axis) and characteristic length $-\zeta=C/(dC/dx)$ (dashed curve, right axis) for the same parameters as in (a).}
\label{fig:capacitor}
\end{center}
\end{figure}

\section{Capacitor}
 We provide a specific design which admits a strong coupling of a nanomechanical resonator to an LC circuit and simultaneously admits optical coupling  (Fig.~\ref{f:setup}).
 Specifically we envision a parallel plate capacitor 
where we replace one of the plates by a set of wires of thickness $t$, width $r$, and separated by a distance $d$, see Fig.~\ref{fig:capacitor}(a). In order to have an interaction with light we introduce a hole in the capacitor to allow for laser beams to get through (not shown).  This hole has a minimal effect on the capacitance when the membrane area is much larger than the mode cross section of the cavity, consistent with a (1 mm)$^2$ membrane and a 50 $\mu$m cavity waist.  A dielectric membrane of thickness $h$ is inserted into the capacitor a distance $x_m$ from the top of the wires. The replacement of  one of the capacitor plates by the set of wires creates a  spatial inhomogeniety of the electric field, which attracts the membrane towards the wires when the capacitor is charged.   At the same time this inhomogeniety also means that the capacitance $C(x_{m})$ will depend on the  position of the membrane $x_{m}$. Expanding the capacitance around the 
 equilibrium position gives rise to 
an LC-membrane coupling $\propto x_m q^2$. This is analogous to the radiation pressure coupling that occurs in the optical domain, i.e., a $\chi^{(2)}$ type nonlinearity.  As such, we can enhance the coupling strength by providing a classical displacement of the LC circuit's charge, with either a dc or ac voltage bias $V$ providing an offset charge, $q_o$. For simplicity we restrict ourselves to the case of a dc voltage, though generalization to the ac case is a simple extension of these ideas and allows to frequency match the LC and mechanical systems. The coupling between the membrane position and the charge fluctuations around the equilibrium $\hat q=q-q_0$ is then $\propto x_m q_0 \hat q$ and is enhanced by the large charge $q_0$ induced on the capacitor, in direct analogy to the similar effect for cavity optomechanics.   The full Hamiltonian including electrical and mechanical contributions is:
\begin{equation}
H = \frac{\phi^2}{2 L} + \frac{p_m^2}{2 m} + \frac{m \omega_m^2 (x_m-x_e)^2}{2} +\frac{q^2}{2C(x_m)} - q V\ .
\end{equation}
 Here $x_e$ is the equilibrium membrane position at $V=0$. 
The fixed point of the classical charge $q_0$ at a given bias voltage and the equilibrium displacement $X$ are then found from $\partial_q H|_{q_0,X} = \partial_{x_{m}} H |_{q_0,X} = 0$ 
which yields
\begin{eqnarray*}
q_0 & = & C(X) V \\
X  & = & x_e - \frac{q^2}{2 C} \frac{1}{\omega_m} \frac{x_0^2}{\zeta} 
\end{eqnarray*}
Here we have introduced a characteristic length scale $\zeta$ defined by $\zeta^{-1}=-1/C\cdot {\partial C}/{\partial x_{m}}|_X$, which describes the relative change in the capacitance at the new equilibrium position $X$.

Around these classical values, we consider the remaining quantum fluctuations $\hat q, \hat x$.  We
 change to annihilation and creation operators $\hat a (\hat b)$ for the membrane (LC), and find a Hamiltonian
\begin{equation}
H = \omega_m \hat a^\dag \hat a + \omega_0 \hat b^\dag \hat b + \frac{g}{2} (\hat a + \hat a^\dag) (\hat b + \hat b^\dag)
\end{equation}
with $g = \sqrt{\omega_m \omega_0} \sqrt{({x_e - X})/{ 2 \zeta}}$. This corresponds to the model Hamiltonian examined in the beginning. 
Assuming a constant value of $\partial C/\partial x_{m}$  from $x_e$ to $X$, the coupling constant can be expressed in a more intuitive form $g=\sqrt{\omega_m \omega_0} \sqrt{\frac{\Delta C}{2C}}$, i.e., through the capacitance change $\Delta C$  caused by the displacement of the membrane due to the applied voltage $V$. 
The solution is a stable point under the condition $g<\sqrt{\omega_0\omega_m}$.
We have neglected a small nonlinear correction, due to the femtometer length scale of the zero-point membrane fluctuations.
Devices which enhance this nonlinear coupling are of interest, but beyond the scope of the present work. 

To obtain quantitative estimates of the feasible coupling constant, we assume the plate electrode to be much larger than the separation of the plate and transverse dimensions of the wires $\sqrt{A}\gg d,D,r$. We can then find the capacitance for a given position of the membrane by 
solving for the potential using the finite element method.
We express the capacitance as $C(x_{m})=\frac{\epsilon_0 A}{D}c(x_{m})$,
where $c(x_{m})$ is a dimensionless number of order unity, which describe the deviation from a standard parallel plate capacitor.

As an example, we take a SiN membrane of dielectric constant $\epsilon=7.6$ and thickness $h=100$ nm inserted into a capacitor with a separation $D=2$ $\mu$m and dimensions $r=D/4$, $d=3D/4$, and $t=D/4$. A simulation with these values is shown in Fig. \ref{fig:capacitor} (b). From this simulation we extract the values of $\zeta\approx 30D=60 $ $\mu$m for a distance of $X\approx 0.2D=0.4$ $\mu$m.
Hence if the applied voltage shifts the equilibrium position by  $x_e-X\approx10$ nm around 
$X\approx 0.2D=0.4$ $\mu$m the coupling constant is  $g/\omega\approx 0.01$ and the system is in the strong coupling regime if the Q values of the LC circuit and membrane exceed 100.  Assuming an operating frequency of $\omega=(2\pi)1$ MHz and an oscillator length $x_0=3$ fm this displacement only requires an applied voltage on the order of a few volts. If a larger initial separation is desirable, a similar coupling ($g/\omega \approx 0.02$) could be achieved if the equilibrium distance is shifted from $x_e\approx 0.6D=1.2 $ $\mu$m to $X\approx 0.4D=0.8$ $\mu$m ($\zeta\approx 400D= 800$ $\mu$m for  $X\approx 0.4D=0.8$ $\mu$m) with an applied voltage of several hundred volts.

\section{Cooling the LC circuit}
The membrane may be efficiently cooled via optomechanical coupling
between the radiation pressure force of a cavity field and the
position of the central area of the membrane.  The details of this
process have been analyzed by a wide variety of groups~\cite{Girvin}.  In essence the effect on  the
membrane degree of freedom $\hat a$ is to induce a
damping $\Gamma_m$, which is much greater than the intrinsic damping rate of the membrane $\Gamma_m \gg \gamma_m$, but is limited by the cavity decay  rate $\Gamma_{m}\lesssim \kappa$. This additional damping only produces  moderate additional
quantum fluctuations associated with the vacuum noise of the light
field (which can be accounted for by adjusting the temperature of $\hat a_{in}$ defined below).  Working with the LC circuit with damping $\gamma$, resonant with the membrane ($\omega_0 =
\omega_m=\omega$), we may expect an efficient coupling between the
optomechanical system and the LC circuit, provided that $g>\gamma$ such that we can get excitations out of the system faster than they leak in. 

We use the input-output formalism in the rotating wave approximation to find a full description of this combined mode cooling.  The Heisenberg-Langevin equations describing this situation are 
\begin{eqnarray*}
  \dot{\hat a} & = & - \Gamma_m \hat a + \sqrt{2 \gamma_m} \hat a_{in} - i \frac{g}{2} \hat b \\
\dot{\hat b} & = & - \gamma \hat b + \sqrt{2 \gamma} \hat b_{in} - i \frac{g}{2} \hat a.
\end{eqnarray*}

In a strong damping limit ($\Gamma_m > g$), we can treat the coupled
LC resonator as a perturbation and arrive at
\begin{equation*}
  \dot{\hat b}  \approx  -( \gamma+\Gamma) \hat b + \sqrt{2 \gamma} \hat b_{in} - i
  \frac{g}{2\Gamma_m} \sqrt{2 \gamma_m} \hat a_{in}. 
  \end{equation*}
This equation describes the cooling of $\hat b$ through the membrane-light system with a rate $\Gamma = 
g^2/4 \Gamma_m$.  In the continuous cooling limit, we
expect to achieve a thermal population in $\hat b$ given by
\begin{equation*}
  \mean{\hat b^\dag \hat b}
\approx \frac{\gamma}{\Gamma+\gamma} n_b + \frac{
    2\gamma_m }{ g} n_a,
\end{equation*}
where $n_a,n_b$ are the original thermal occupation of modes $\hat a$ and $\hat b$. Typically this population will be dominated by the heating of the LC circuit (the first term) since the membrane can have a very large $Q\sim 10^{6}$~\cite{Kimble}.  

In the mode-resolved, strong-coupling limit, with $\omega_0 = \omega_m=\omega$ and $\gamma,\Gamma_m < g$, each of the two normal modes $\hat a + \hat b$ and $\hat a -
\hat b$ have frequencies $\omega \pm g/2$ and a damping rate given given by the average of the two damping rates $(\gamma+\Gamma_{m})/2$. The optomechanical
coupling then works independently on each of the two modes. Assuming  again the $Q$ of the membrane to be much higher than the $Q$ of the LC, a standard argument for optomechanical cooling~\cite{marquardt08} leads to a thermal occupation number
 of ${\gamma}n_b/{\Gamma_{m}}$.  
 
Comparing the two limits derived above we see that the minimal thermal occupation is achieved at a cooling laser power and detuning such that $\Gamma_{m}\approx g$, where we obtain a population $\gamma n_{b}/g$. The cooling of the membrane is, however, limited by the cavity decay rate $\Gamma_{m}\lesssim \kappa$. Hence the cooling limit is the larger of $\gamma n_{b}/g$ and $\gamma n_{b}/\kappa$.  We have neglected optical heating effects, consistent with our  assumption of good sideband-resolution ($\kappa < \omega$).

\section{Sensitivity of optical readout of LC circuit}
The cooling identified above  realizes an interface between optical fields and rf excitations of the LC circuits at the single photon level, and we now turn to a possible application of this interface. Often 
 LC circuits are used in sensitive detectors to pick up very small signals~\cite{NMRloop}.
As we will now show, the sensitivity in such experiments can be improved by detecting the cooling light leaving the cavity.  This takes advantage of the fact that the homodyne detection of laser light can be quantum noise limited with near-unit quantum efficiency, thus avoiding many of the noise sources present for low frequency signals. To show this we will work in the strong damping limit identified above $\Gamma_{m}> g$ with the LC circuit  tuned into resonance with the membrane ($\omega_{0}=\omega_{m}=\omega$). Again we also assume the damping of the mechanical motion of the membrane to be negligible $\gamma_{m}\ll g$.  In this limit  the membrane and the cavity mediate an effective interaction between the LC mode $\hat b$ and the optical cavity input/output modes $\hat d$ with the effective cooling  rate $\Gamma$  introduced above. In the rotating wave approximation this situation is described by the generic equations
\begin{eqnarray}
\dot{\hat b}&=&-(\gamma+\Gamma) \hat b+if(t)+\sqrt{2\gamma}\hat b_{in}-\sqrt{2\Gamma} \hat c_{in}\\
\hat{d}_{out}&=&\hat d_{in}+\sqrt{2\Gamma}b.
\end{eqnarray}
Here, we have introduced an incoming signal to be measured, which  is described by $f(t)$.  If the signal is from a voltage $V$ applied to the system, $f(t)=-\left(C/4\hbar^2L\right)^{1/4} V(t)$.

Suppose that we  want to measure the Fourier components of the incoming signals $f(\nu)$ detuned by a frequency $\nu$ with respect to the resonance frequency of the LC circuit $\omega_{0}$ within a certain  bandwidth $|\nu|\lesssim \delta \omega_0$. This can be done by splitting the outgoing signal on a beamsplitter ($d_{\pm}=(d\pm d_{b})/\sqrt{2}$, where $d_{b}$  is the annihilation operator for the other mode incident on the beamsplitter) and inferring the two quadratures  $x_f(\nu)=(f(\nu)+f^*(\nu))/\sqrt{2}$ and $p_{f}=(f(\nu)-f^*(\nu))/i\sqrt{2}$ from a homodyne detection of the $x_{+}=(d_{+}+d_{+}^\dagger)/\sqrt{2}$ ($p_{-}=(d_{-}-d_{-}^\dagger)/i\sqrt{2}$) quadrature of the $d_{+}$ ($d_{-}$) mode. The signal-to-noise ratio for, e.g., a measurement of the amplitude can be defined by $S=(\langle x_+(\nu)^{2}\rangle+\langle p_-(\nu)^{2})/2N$. Here  $N$ describes the noise $\langle x_{+}(\nu)x_{+}(\nu')\rangle=\langle p_{-}(\nu)p_{-}(\nu')\rangle =N\delta(\nu-\nu'
 )$ in the absence of any signal. From the equations of motion we find
\begin{equation}
S=\frac{2\Gamma|\langle f(\nu)\rangle|^2 }{{2\gamma\Gamma (2n_b+1) +{\left[\gamma^2+\Gamma^2+\nu^2\right]}(2n_d+1)}}
\end{equation}
Here 
$n_d$ describes the number of thermal excitations in 
the field used to probe the circuit 
and we assume that the fields incident on the beamsplitter are of the same type such that $\langle d^\dagger d\rangle=\langle d_{b}^{\dagger} d_{b}\rangle$. 

Let us compare our approach to the case where the LC circuit is read out by homodyne detection with an rf amplifier assumed to have a similar number of thermal excitations as the system being measured $n_d=n_b\gg 1$. Disregarding any additional noise added during the amplification, $S$ is optimized for $\Gamma=\gamma$ and is  limited to $S=|\langle f(\nu)\rangle|^{2}/4\gamma n_b$ with a detection bandwidth $\delta \omega=2\gamma$.  In contrast, with the optical readout, the incoming laser fields can be quantum noise limited with $n_d=0$ if light is in a coherent state. In this case we obtain twice the signal-to-noise ratio $S$ for $\gamma\leq \Gamma\lesssim \gamma n_b$. The optimal sensitivity is thus  better with optical detection, even if we assume ideal detection of the fields in both cases. Such an ideal detection is routinely achieved by homodyne detection of optical fields with near unity quantum efficiency, whereas it is hard to achieve for rf fields. For realistic limited detect
 or efficiencies of rf fields, the sensitivity may thus be significantly improved using optical readout. Furthermore the high sensitivity with laser cooling is obtained over a much larger bandwidth which is determined by $\delta \omega=2 \sqrt{2\Gamma \gamma n_b}$. In other words, if, prior to laser cooling, the LC circuit had a high-Q and a narrow bandwidth $\gamma$ that is less than the bandwidth $\delta \omega_0$ required for a particular application, laser cooling allows an increase of the bandwidth with a limited decrease in the sensitivity ($< 3$ dB) if $\delta \omega_0 \lesssim \delta\omega$.  Using regular rf techniques, an alternative approach would be to increase the bandwidth by
 increasing the damping of the circuit, but this would result in a decrease of the sensitivity by a factor of  $\sqrt{\delta \omega_0 / \gamma}$. Crucially, since optical fields are shot-noise limited even at room temperature, this measurement setup does not involve cryogenics. 
 
The potential benefits of this approach -- high quantum efficiency conversion from radio frequency to optical photons, and the corresponding potential for low temperature detection of radio frequency signals -- are limited by the finite Q values for room temperature inductors.  Appropriate replacements may be considered, such as crystal resonators or cryogenic superconducting resonators.  An additional benefit of a cryogenic setup is the possibility to enter the quantum strong coupling limit, $g \gtrsim \gamma n_{\rm thermal}$, at which point the conversion from radio frequency to optical domain can be used as a quantum interface.
However, understanding of these features and improvements requires further investigation.

We gratefully acknowledge useful discussions with Koji Usami, Ole Hansen, Silvan Schmidt, Anja Boisen, and John Lawall.  JMT thanks the NBI for hospitality during his stay.  This research was funded by ARO MURI award W911NF0910406, DARPA and by the EU project Q-ESSENCE.

\end{document}